\documentclass[twocolumn,showpacs,prb]{revtex4}
\usepackage{amssymb}
\usepackage{makeidx}
\usepackage{amsmath}
\usepackage{graphicx}
\usepackage{epstopdf}
\usepackage{dcolumn}
\usepackage{bm}

\begin{document}

\title{Coherent nonlinear optical response of graphene in the quantum Hall
regime }
\author{H.K. Avetissian}
\author{G.F. Mkrtchian}
\affiliation{Centre of Strong Fields Physics, Yerevan State University, 1 A. Manukian,
Yerevan 0025, Armenia}

\begin{abstract}
We study nonlinear optical response of graphene in the quantum Hall regime
to an intense laser pulse. In particular, we consider harmonic generation
process. We demonstrate that the generalized magneto-optical conductivity of
graphene on the harmonics of a strong pump laser radiation has a
characteristic Hall plateaus feature. The plateaus heights depend on the
laser intensity and broadening of Landau levels, so that are not quantized
exactly. This nonlinear effect remains robust against the significant
broadening of Landau levels. We predict realization of an experiment through
the observation of the third harmonic signal and nonlinear Faraday effect,
which are within the experimental feasibility.
\end{abstract}

\maketitle



\section{INTRODUCTION}

Graphene, a single sheet of carbon atoms in a honeycomb 2D lattice, has
attracted a tremendous interest since its experimental realization.\cite%
{Nov1,Nov2} Graphene possesses several exotic features because of a linear
energy dispersion law around the two nodal points $K$ and $K^{\prime }$ in
the Brillouin zone. Due to its unique electronic band structure and specific
electromagnetic properties (in particular, high degree of nonlinearity\cite%
{Mer1-2}), graphene is widely considered as a promising material in various
nonlinear optical applications. Hence, it is of actual interest to study
graphene fundamental physics in the presence of an electric, magnetic,
and/or optical fields. When a static magnetic field (uniform) is applied
perpendicular to the graphene plane, the electron energy is quantized
forming nonequidistant Landau levels (LL). For the massless Dirac
quasiparticles extra LL appears with zero energy shared by both electrons
and holes without an energy gap. As a consequence, in the graphene the
anomalous half-integer quantum Hall effect (QHE) takes place.\cite%
{Nov3,Zhang,GS,Goer} This differs strongly from a integer QHE in
conventional two-dimensional electron gas.\cite{QHE} As is well known, in
the static QHE the Hall resistance is quantized into plateaus, which is a
topological phenomenon. \cite{TKNN} Thus, being topologically protected it
is robust against details of the disorder. Topological analysis of the QHE
in graphene\cite{GTop} shoes that the zero-mass Dirac QHE persists up to the
van Hove singularities. Optical response of the\ graphene QHE system has
been studied by different groups. Optical conductivity of graphene QHE
system with the different aspects have been studied in Refs. [%
\onlinecite{MHA,AC1,AC2,AC3,AC4,LLL1,Ferreira}]. Ultrafast carrier dynamics
and carrier multiplication in Landau-quantized graphene have been
investigated in Refs. [\onlinecite{Carrier1,Carrier2}]. In Refs. [%
\onlinecite{LLL1,LLL2}] tunable graphene-based laser on the Landau levels in
the terahertz regime have been proposed. As was shown in Ref. [%
\onlinecite{MHA}] the plateau structure in the QHE in 2DEG and in graphene
is retained, up to significant degree of disorder, even in the ac (THz)
regime, although the heights of the plateaus are no longer quantized. While
most of the works have concentrated on static or linear optical properties
of the graphene QHE system, one direction that has not been fully explored
is the nonlinear response of Landau-quantized graphene to a strong laser
radiation, which is the purpose of the present study. Note that in the
absence of the magnetic field graphene is an effective material for
multiphoton interband excitation, wave mixing, and harmonic generation.\cite%
{Mer1-2,Mer3-4} Thus, in the magnetic field due to peaks in the density of
states one can expect enhancement of the harmonic radiation power.

In the graphene QHE system, wave-particle interaction can be characterized
by the dimensionless parameter $\chi =eE_{0}l_{B}/\hbar \omega $, which
represents the work of the wave electric field $E_{0}$ on the magnetic
length $l_{B}=\sqrt{c\hbar /eB}$ ($e$ is the elementary charge, $\hbar $ is
Planck's constant, $c$ is the light speed in vacuum, and $B$ is the magnetic
field strength) in units of photon energy $\hbar \omega $. Depending on the
value of this parameter $\chi $, one can distinguish three different regimes
in the wave-particle interaction process. Thus, $\chi <<1$ corresponds to
the one-photon interaction regime, $\chi >>1$ corresponds to the static
field limit, and $\chi \sim 1$ to the multiphoton interaction regime. In
this paper we consider just multiphoton interaction regime and look for
features in the harmonic spectra of the laser driven graphene. Accordingly,
the time evolution of the considered system is found using a nonperturbative
numerical approach, revealing that the generalized magneto-optical
conductivity of graphene on the harmonics of a strong pump laser radiation
in the quantum Hall regime has a characteristic Hall plateaus feature. The
plateaus heights are not quantized exactly and depend on the laser intensity
and broadening of Landau levels. The effect remains robust against a
significant broadening of Landau levels and takes place for wide range of
intensities and frequencies of a pump wave.

The paper is organized as follows. In Sec. II the Hamiltonian which governs
the quantum dynamics of considered process and the set of equations for a
single-particle density matrix are formulated. In Sec. III, we numerically
solve obtained equations and consider coherent nonlinear optical response of
graphene in the quantum Hall regime on the fundamental harmonic with
limiting linear case and on the third harmonic of an incident wave. Finally,
conclusions are given in Sec. IV.

\section{BASIC MODEL AND EVOLUTIONARY EQUATION\ FOR DENSITY MATRIX}

We begin our study with construction of the Hamiltonian which governs the
quantum dynamics of considered process. The graphene sheet is taken in the $%
xy$ plane ($z=0$) and a uniform static magnetic field is applied in the
perpendicular direction. A plane linearly polarized (along the $x$ axis)
quasimonochromatic electromagnetic radiation of carrier frequency $\omega $
and slowly varying envelope $E_{0}(t)$ interacts with the such system. To
avoid nonphysical effects semi-infinite pulses with smooth turn-on, in
particular, with hyperbolic tangent $E_{0}(t)=E_{0}\mathrm{tanh}(t/\tau
_{r}) $ envelope is considered. Here the characteristic rise time $\tau _{r}$
is chosen to be $\tau _{r}=20\pi /\omega $. We consider the case of
interaction when the wave propagates in the perpendicular direction to the
graphene sheet to exclude the effect of the wave magnetic field.\cite{Mer1-2}
Under these circumstances the single-particle Hamiltonian of graphene QHE
system in the presence of a uniform time-dependent electric field 
\begin{equation}
E(t)=E_{0}(t)\cos \omega t  \label{e1}
\end{equation}%
reads: 
\begin{equation}
\mathcal{H}_{s}=\hbar \omega _{B}\left( 
\begin{array}{cc}
0 & \widehat{a} \\ 
\widehat{a}^{\dagger } & 0%
\end{array}%
\right) +\left[ \widehat{I}\frac{el_{B}E(t)}{\sqrt{2}}\left( \widehat{b}+i%
\widehat{a}\right) +\mathrm{h.c.}\right] .  \label{ham}
\end{equation}%
Here $\omega _{B}=\sqrt{2}\mathrm{v}_{F}/l_{B}$ plays the role of the
cyclotron frequency, $\mathrm{v}_{F}$ is the Fermi velocity, and $\widehat{I}
$ is the identity matrix. For the interaction Hamiltonian we use a length
gauge describing the interaction by the potential energy. The ladder
operators $\widehat{a}$ and $\widehat{a}^{\dagger }$ describe quantum
cyclotron motion, while $\widehat{b}$ and $\widehat{b}^{\dagger }$
correspond to guiding center motion. These ladder operators satisfy the
usual commutation relations $[\widehat{a},\widehat{a}^{\dagger }]=1$ and $[%
\widehat{b},\widehat{b}^{\dagger }]=1$. The Hamiltonian is applicable near
the $K$ point in the Brillouin zone. For the considered model there are no
intervalley transitions and the valley index, as well as spin index will be
skipped. They will be incorporated into the consideration via the spin $%
g_{s} $ and valley $g_{v}$ degeneracy factors.

The single free particle Hamiltonian, that is the first term in Eq. (\ref%
{ham}) can be diagonalized analytically.\cite{Goer} The wave function and
energy spectrum are given by 
\begin{equation}
|\psi _{n,m}\rangle =\sqrt{\left( 1+\delta _{n,0}\right) /2}\left( 
\begin{array}{c}
\mathrm{sgn}(n)|\left\vert n\right\vert -1,m\rangle \\ 
|\left\vert n\right\vert ,m\rangle%
\end{array}%
\right) ,  \label{WF}
\end{equation}%
\begin{equation}
\varepsilon _{n}=\mathrm{sgn}(n)\hbar \omega _{B}\sqrt{\left\vert
n\right\vert }.  \label{energy}
\end{equation}%
Here $|\left\vert n\right\vert ,m\rangle \ =|\left\vert n\right\vert \rangle
\ \otimes |m\rangle $, with $|\left\vert n\right\vert \rangle $ and $%
|m\rangle $ being the harmonic oscillator wave functions. The eigenstates (%
\ref{WF}) are defined by the quantum numbers $n=0,\pm 1...$ and $m=0,1..$.
Here $n$ is the LL index -- for an electron $n>0$ and for a hole $n<0$. The
extra LL with $n=0$ is shared by both electrons and holes. The LLs are
degenerate upon second quantum number $m$ with the large degeneracy factor $%
N_{B}=\mathcal{S}/2\pi l_{B}^{2}$ which equals the number of flux quanta
threading the 2D surface $\mathcal{S}$ occupied by the electrons. The terms $%
\sim \widehat{a}E(t)$ in the Hamiltonian (\ref{ham}) describe transitions
between LLs, while the terms $\sim \widehat{b}E(t)$ describe transitions
within the same LL. These transitions can be excluded from the consideration
by the appropriate choice of Dirac states for the construction of the
carrier quantum field operators. Expanding the fermionic field operator 
\begin{equation}
|\widehat{\Psi }\rangle =\sum\limits_{n,m}\widehat{a}_{n,m}|\widetilde{\psi }%
_{n,m}\rangle  \label{expand}
\end{equation}%
over the dressed states%
\begin{equation}
|\widetilde{\psi }_{n,m}\rangle =\exp \left[ -\frac{i}{\hbar }\frac{el_{B}}{%
\sqrt{2}}\int_{0}^{t}E(t^{\prime })dt^{\prime }\left( \widehat{b}^{\dagger }+%
\widehat{b}\right) \right] |\psi _{n,m}\rangle ,  \label{free}
\end{equation}%
the Hamiltonian of the system in the second quantization formalism 
\begin{equation*}
\widehat{H}=\left\langle \widehat{\Psi }\right\vert \mathcal{H}%
_{s}\left\vert \widehat{\Psi }\right\rangle
\end{equation*}%
can be presented in the form:%
\begin{eqnarray}
\widehat{H} &=&\sum\limits_{n=-\infty }^{\infty
}\sum\limits_{m=0}^{N_{B}-1}\varepsilon _{n}\widehat{a}_{n,m}^{+}\widehat{a}%
_{n,m}  \notag \\
&&+\sum\limits_{n,n^{\prime }=-\infty }^{\infty
}\sum\limits_{m=0}^{N_{B}-1}E(t)\mathcal{D}_{n,n^{\prime }}\widehat{a}%
_{n,m}^{+}\widehat{a}_{n^{\prime },m},  \label{gr2}
\end{eqnarray}%
where $\widehat{a}_{n,m}^{\dagger }$\ and $\widehat{a}_{n,m}$ are,
respectively, the creation and annihilation operators for a carrier in a LL
state, and $\mathcal{D}_{n,n^{\prime }}$ is the dipole moment operator: 
\begin{equation}
\mathcal{D}_{n,n^{\prime }}=\frac{iel_{B}}{2\sqrt{2}}\left[ \varkappa
_{n,n^{\prime }}\mathbb{\delta }_{\left\vert n\right\vert ,\left\vert
n^{\prime }\right\vert +1}+\varkappa _{n^{\prime },n}\mathbb{\delta }%
_{\left\vert n\right\vert ,\left\vert n^{\prime }\right\vert -1}\right] 
\frac{\hbar \omega _{B}}{\varepsilon _{n^{\prime }}-\varepsilon _{n}},
\label{Dnn'}
\end{equation}%
where $\varkappa _{n,n^{\prime }}=\mathrm{sgn}(n)\sqrt{1+\delta _{n^{\prime
},0}}$. Then we will pass to Heisenberg representation where operators obey
the evolution equation 
\begin{equation*}
i\hbar \frac{\partial \widehat{L}}{\partial t}=\left[ \widehat{L},\widehat{H}%
\right]
\end{equation*}%
and expectation values are determined by the initial density matrix $%
\widehat{D}$: $<\widehat{L}>=Sp\left( \widehat{D}\widehat{L}\right) $. In
order to develop microscopic theory of the nonlinear interaction of the
graphene QHE system with a strong radiation field, we need to solve the
Liouville-von Neumann equation for the single-particle density matrix%
\begin{equation}
\rho (n_{1},m_{1};n_{2},m_{2},t)=<\widehat{a}_{n_{2},m_{2}}^{+}(t)\widehat{a}%
_{n_{1},m_{1}}(t)>.  \label{grSPDM}
\end{equation}%
For the initial state of the graphene quasiparticles we assume an ideal
Fermi gas in equilibrium. According to the latter, the initial
single-particle density matrix will be diagonal, and we will have the
Fermi-Dirac distribution:%
\begin{equation}
\rho (n_{1},m_{1};n_{2},m_{2},0)=\rho _{F}\left( n_{1}\right) \frac{\delta
_{n_{1},n_{2}}\delta _{m_{1},m_{2}}}{1+\exp \left( \frac{\varepsilon
_{n_{1}}-\varepsilon _{F}}{T}\right) },  \label{grISPDM}
\end{equation}%
\begin{equation}
\rho _{F}\left( n_{1}\right) =\frac{1}{1+\exp \left( \frac{\varepsilon
_{n_{1}}-\varepsilon _{F}}{T}\right) }.  \label{FDD}
\end{equation}%
Including in Eq. (\ref{grISPDM}) quantity $\varepsilon _{F}$ is the Fermi
energy, $T$ is the temperature in energy units. As is seen from the
interaction term in the Hamiltonian (\ref{gr2}) quantum number $m$ is
conserved: $\rho (n_{1},m_{1};n_{2},m_{2},t)=\rho _{n_{1},n_{2}}\left(
t\right) \delta _{m_{1},m_{2}}$. To include the effect of the LLs broadening
we will assume that it is caused by the disorder described by randomly
placed scatterers. When the range of the random potential is larger than the
lattice constant in graphene, the scattering between $K$ and $K^{\prime }$
points in the Brillouin zone is suppressed and we can assume homogeneous
broadening of the LLs.\cite{Ando} The latter can be incorporated into
evolution equation for $\rho _{n_{1},n_{2}}\left( t\right) $ by the damping
term $-i\Gamma _{n_{1},n_{2}}\rho _{n_{1},n_{2}}\left( t\right) $ and from
Heisenberg equation one can obtain evolution equation for the reduced
single-particle density matrix:%
\begin{equation}
i\hbar \frac{\partial \rho _{n_{1},n_{2}}(t)}{\partial t}=\left[ \varepsilon
_{n_{1}}-\varepsilon _{n_{2}}\right] \rho _{n_{1},n_{2}}(t)-i\Gamma
_{n_{1},n_{2}}\rho _{n_{1},n_{2}}\left( t\right)  \notag
\end{equation}%
\begin{equation}
-E(t)\sum\limits_{n}\left[ \mathcal{D}_{n,n_{2}}\rho _{n_{1},n}(t)-\mathcal{D%
}_{n_{1},n}\rho _{n,n_{2}}(t)\right] .  \label{grevol}
\end{equation}%
For the norm-conserving damping matrix we take $\Gamma _{n_{1},n_{2}}=\Gamma
\left( 1-\delta _{n_{1},n_{2}}\right) $, where $\Gamma $ measures the Landau
level broadening.

\section{COHERENT NONLINEAR OPTICAL RESPONSE}

Solving Eq. (\ref{grevol}) with the initial condition (\ref{grISPDM}) one
can reveal nonlinear response of the graphene QHE system to a strong laser
radiation. At that one can expect intense radiation of harmonics of the
incoming wave-field in the result of the coherent transitions between LLs.
The harmonics will be described by the additional generated fields $%
E_{x,y}^{(g)}$. \ We assume that the generated fields are considerably
smaller than the incoming field $\left\vert E_{x,y}^{(g)}\right\vert
<<\left\vert E\right\vert $. In this case we do not need to solve
self-consistent Maxwell's wave equation with Heisenberg equations. To
determine the electromagnetic field of harmonics we can solve Maxwell's wave
equation in the propagation direction with the given source term:%
\begin{equation}
\frac{\partial ^{2}E_{x,y}^{(t)}}{\partial z^{2}}-\frac{1}{c^{2}}\frac{%
\partial ^{2}E_{x,y}^{(t)}}{\partial t^{2}}=\frac{4\pi }{c^{2}}\frac{%
\partial \mathcal{J}_{x,y}\left( t\right) }{\partial t}\delta \left(
z\right) .  \label{Max}
\end{equation}%
Here $\delta \left( z\right) $ is the Dirac delta function, $E_{x,y}^{(t)}$
is the total field, and $\mathcal{J}_{x,y}\left( t\right) $ is the mean
value of the surface current density operator: 
\begin{eqnarray}
\widehat{\mathcal{J}}_{x}\left( t\right) &=&-\frac{e\mathrm{v}_{F}g_{s}g_{v}%
}{\mathcal{S}}\left\langle \widehat{\Psi }\right\vert \sigma _{x}\left\vert 
\widehat{\Psi }\right\rangle ,  \notag \\
\widehat{\mathcal{J}}_{y}\left( t\right) &=&-\frac{e\mathrm{v}_{F}g_{s}g_{v}%
}{\mathcal{S}}\left\langle \widehat{\Psi }\right\vert \sigma _{y}\left\vert 
\widehat{\Psi }\right\rangle .  \label{jqxy}
\end{eqnarray}%
Here $g_{s}=2$\ and $g_{v}=2$\ are the spin and valley degeneracy factors,
respectively. With the help of Eqs. (\ref{expand}) and (\ref{grSPDM}), the
expectation value (\ref{jqxy}) of the total current in components can be
written in the following form:%
\begin{equation*}
\mathcal{J}_{x}=-\frac{e\mathrm{v}_{F}}{\pi l_{B}^{2}}\sum\limits_{n,n^{%
\prime }}\rho _{n^{\prime },n}\left( \varkappa _{n,n^{\prime }}\mathbb{%
\delta }_{\left\vert n\right\vert ,\left\vert n^{\prime }\right\vert
+1}+\varkappa _{n^{\prime },n}\mathbb{\delta }_{\left\vert n\right\vert
,\left\vert n^{\prime }\right\vert -1}\right) ,
\end{equation*}%
\begin{equation}
\mathcal{J}_{y}=-\frac{ie\mathrm{v}_{F}}{\pi l_{B}^{2}}\sum\limits_{n,n^{%
\prime }}\rho _{n^{\prime },n}\left( \varkappa _{n^{\prime },n}\mathbb{%
\delta }_{\left\vert n\right\vert ,\left\vert n^{\prime }\right\vert
-1}-\varkappa _{n,n^{\prime }}\mathbb{\delta }_{\left\vert n\right\vert
,\left\vert n^{\prime }\right\vert +1}\right) .  \label{grcurry}
\end{equation}%
In Eq. (\ref{grcurry}) we have also made summation over the quantum number $%
m $ which yields the degeneracy factor $\mathcal{S}/2\pi l_{B}^{2}$. The
solution to equation (\ref{Max}) reads%
\begin{equation*}
E_{x,y}^{(t)}\left( t,z\right) =E_{x,y}\left( t-z/c\right)
\end{equation*}%
\begin{equation}
-\frac{2\pi }{c}\left[ \theta \left( z\right) \mathcal{J}_{x,y}\left(
t-z/c\right) +\theta \left( -z\right) \mathcal{J}_{x,y}\left( t+z/c\right) %
\right] ,  \label{sol}
\end{equation}%
where $\theta \left( z\right) $ is the Heaviside step function with $\theta
\left( z\right) =1$ for $z\geq 0$ and zero elsewhere. The first term in Eq. (%
\ref{sol}) is the incoming wave. In the second line of Eq. (\ref{sol}), we
see that after the encounter with the graphene sheet two propagating waves
are generated. One traveling in the propagation direction of the incoming
pulse and one traveling in the opposite direction. The Heaviside functions
ensure that the generated light propagates from the source located at $z=0$.
We assume that the spectrum is measured at a fixed observation point in the
forward propagation direction. For the generated field at $z>0$ we have%
\begin{equation}
E_{x,y}^{(g)}\left( t-z/c\right) =-\frac{2\pi }{c}\mathcal{J}_{x,y}\left(
t-z/c\right) .  \label{solut}
\end{equation}

Now, performing the summation in Eqs. (\ref{grcurry}) and using solutions (%
\ref{solut}) we can calculate the harmonic radiation spectrum with the help
of Fourier transform of the functions $E_{x,y}^{(g)}\left( t-z/c\right) $:%
\begin{equation}
E_{x,y}^{(g)}\left( s\right) =\frac{\omega }{2\pi }\int_{0}^{2\pi /\omega
}E_{x,y}^{(g)}\left( t\right) e^{is\omega t}dt=-\frac{2\pi }{c}\mathcal{J}%
_{x,y}^{\left( s\right) },  \label{F1}
\end{equation}%
where%
\begin{equation}
\mathcal{J}_{x,y}^{\left( s\right) }=\frac{\omega }{2\pi }\int_{0}^{2\pi
/\omega }\mathcal{J}_{x,y}\left( t\right) e^{is\omega t}dt.  \label{F2}
\end{equation}%
Let us now introduce generalized magneto-optical conductivity of graphene 
\begin{equation}
\Sigma _{xx}^{(s)}\left( \omega ,E_{0}\right) =\frac{\mathcal{J}_{x}^{\left(
s\right) }}{E_{x}^{\left( 1\right) }}=\frac{2\mathcal{J}_{x}^{\left(
s\right) }}{E_{0}},  \label{long}
\end{equation}%
\begin{equation}
\Sigma _{yx}^{(s)}\left( \omega ,E_{0}\right) =\frac{\mathcal{J}_{y}^{\left(
s\right) }}{E_{x}^{\left( 1\right) }}=\frac{2\mathcal{J}_{y}^{\left(
s\right) }}{E_{0}}.  \label{hall}
\end{equation}%
Note that in the limit $E_{0}\rightarrow 0$ at $s=1$, from the generalized
conductivity (\ref{long}) and (\ref{hall}) one can derive magneto-optical
conductivity of graphene obtained within the framework of linear response
theory:%
\begin{equation}
\Sigma _{xx}^{(1)}\left( \omega ,E_{0}\rightarrow 0\right) =\sigma
_{xx}\left( \omega \right) ,  \label{limit1}
\end{equation}%
\begin{equation}
\Sigma _{yx}^{(1)}\left( \omega ,E_{0}\rightarrow 0\right) =\sigma
_{yx}\left( \omega \right) .  \label{limit2}
\end{equation}%
For $s\neq 1$ the quantities defined via Eqs. (\ref{long}) and (\ref{hall})
describe nonlinear response of the graphene QHE via radiation of harmonics.
In the following the nonlinear effects on the fundamental harmonic with
limiting linear case and on the third harmonic will be considered separately.

\subsection{Linear magneto-optical conductivity of graphene}

For the comparison with the nonlinear case and for completeness in this
subsection we consider magneto-optical conductivity of graphene in the scope
of linear response theory. Thus, we first solve Eq. (\ref{grevol}) in the
first order over the field $E(t)$. We look for the solution of Eq. (\ref%
{grevol}) in the form%
\begin{equation*}
\rho _{n_{1},n_{2}}(t)\simeq \rho _{n_{1},n_{2}}^{(0)}+\rho
_{n_{1},n_{2}}^{(1)},
\end{equation*}%
with%
\begin{equation*}
\rho _{n_{1},n_{2}}^{(0)}(t)=\rho _{F}\left( n_{1}\right) \delta
_{n_{1},n_{2}}.
\end{equation*}%
Keeping only first order over $E(t)$ terms from Eq. (\ref{grevol}) one can
obtain the solution%
\begin{widetext}
\begin{equation}
\rho _{n_{1},n_{2}}^{(1)}(t)=\frac{\mathcal{D}_{n_{1},n_{2}}\left[ \rho
_{F}(n_{1})-\rho _{F}(n_{2})\right] }{\varepsilon _{n_{1}}-\varepsilon
_{n_{2}}+\hbar \omega -i\Gamma }\frac{E_{0}e^{i\omega t}}{2}+\frac{\mathcal{D%
}_{n_{1},n_{2}}\left[ \rho _{F}(n_{1})-\rho _{F}(n_{2})\right] }{\varepsilon
_{n_{1}}-\varepsilon _{n_{2}}-\hbar \omega -i\Gamma }\frac{E_{0}e^{-i\omega
t}}{2},  \label{LRT}
\end{equation}%
where we have assumed adiabatic turn on of the field $E_{0}(t\rightarrow 0)=0
$. The magneto-optical conductivity of graphene corresponding to obtained
density matrix (\ref{LRT}), according to Eqs. (\ref{long})-(\ref{limit2})
will be given by formulas: 
\begin{equation}
\sigma _{xx}\left( \omega \right) =-\frac{e^{2}}{h}\frac{i}{2}%
\sum\limits_{n_{1},n_{2}}\frac{\rho _{F}(n_{1})-\rho _{F}(n_{2})}{%
\varepsilon _{n_{1}}-\varepsilon _{n_{2}}+\hbar \omega -i\Gamma }\left[
\left( 1+\delta _{n_{2},0}\right) \mathbb{\delta }_{\left\vert
n_{1}\right\vert ,\left\vert n_{2}\right\vert +1}+\left( 1+\delta
_{n_{1},0}\right) \mathbb{\delta }_{\left\vert n_{1}\right\vert ,\left\vert
n_{2}\right\vert -1}\right] \frac{\hbar ^{2}\omega _{B}^{2}}{\varepsilon
_{n_{2}}-\varepsilon _{n_{1}}},  \label{sxx}
\end{equation}%
\begin{equation}
\sigma _{yx}\left( \omega \right) =\frac{e^{2}}{h}\frac{1}{2}%
\sum\limits_{n_{1},n_{2}}\frac{\rho _{F}(n_{1})-\rho _{F}(n_{2})}{%
\varepsilon _{n_{1}}-\varepsilon _{n_{2}}+\hbar \omega -i\Gamma }\left[
\left( 1+\delta _{n_{2},0}\right) \mathbb{\delta }_{\left\vert
n_{1}\right\vert ,\left\vert n_{2}\right\vert +1}-\left( 1+\delta
_{n_{1},0}\right) \mathbb{\delta }_{\left\vert n_{1}\right\vert ,\left\vert
n_{2}\right\vert -1}\right] \frac{\hbar ^{2}\omega _{B}^{2}}{\varepsilon
_{n_{2}}-\varepsilon _{n_{1}}},  \label{syx}
\end{equation}%
\end{widetext}where $h=2\pi \hbar $. Note that Eqs. (\ref{sxx}) and (\ref%
{syx}) for longitudinal ($\sigma _{xx}\left( \omega \right) $) and Hall ($%
\sigma _{yx}\left( \omega \right) $) conductivities coincide with results of
Ref. [\onlinecite{AC2}] obtained via a Green's function calculation and,
also, with a Kubo formula calculation within the Dirac cone approximation.%
\cite{Ferreira} The variation of conductivity with the Fermi energy reveals
characteristic feature of 2D systems: Hall quantization - when Fermi energy
passes through Landau levels. The Hall conductivity quantization rule for
graphene can be obtained from Eq. (\ref{syx}). In the limit of $\left\{
\hbar \omega ,\Gamma ,T\right\} <<\hbar \omega _{B}$ the expression (\ref%
{syx}) simplifies to 
\begin{equation}
\sigma _{yx}\left( 0\right) =4\frac{e^{2}}{h}\left( n_{F}+\frac{1}{2}\right)
,  \label{IQHE}
\end{equation}%
the so-called half-integer quantum Hall quantization rule for graphene\cite%
{Nov3} ($n_{F}$ is the index of the last occupied LL). In Fig. 1 we display
linear optical response of graphene QHE system via real part of Hall
conductivity (\ref{syx}) versus Fermi energy for various frequencies. The LL
broadening is taken to be $\Gamma =0.1\hbar \omega _{B}$. As is seen from
this figure the Hall conductivity quantization rule persist for low
frequency radiation,\cite{MHA,Ferreira} however for higher frequencies it
disappears.

\begin{figure}[tbp]
\includegraphics[width=.46\textwidth]{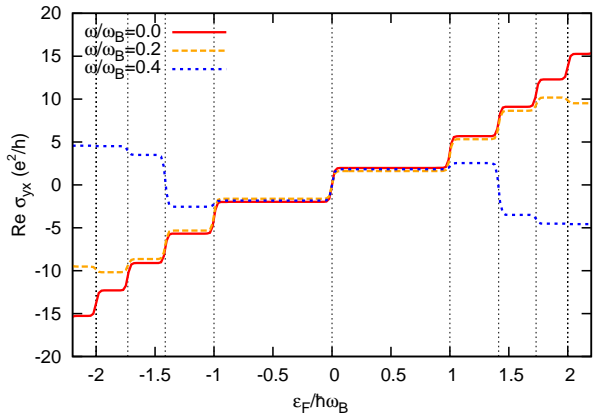}
\caption{Linear optical response of graphene QHE system via real part of the
Hall conductivity versus Fermi energy for various frequencies. The LL
broadening is taken to be $\Gamma =0.1\hbar \protect\omega _{B}$. Vertical
dashed lines indicate Landau levels.}
\end{figure}

\subsection{Non-linear magneto-optical conductivity of graphene}

For the strong fields Eq. (\ref{grevol}) can not be solved analytically and
one should use numerical methods. For this propose the time evolution of
system (\ref{grevol}) is found with the help of the standard fourth-order
Runge-Kutta algorithm and for calculation of Fourier transform of the
functions $\mathcal{J}_{x,y}\left( t\right) $ the fast Fourier transform
algorithm is used. For all calculations the temperature is taken to be $%
T/\hbar \omega _{B}=0.01$.

Figures 2 and 3 show nonlinear response of the graphene QHE system via real
and absolute value of non-linear optical Hall conductivity versus Fermi
energy for various pump wave intensities. From these figures we immediately
notice a step-like structure of the non-linear optical Hall conductivity as
a function of $\varepsilon _{F}$ for various pump wave intensities. In these
figures the vertical dashed lines indicate Landau levels. Although the step
heights are not quantized exactly, the flatness, which is a intrinsic
property of the static QHE, surprisingly exists also in the nonlinear
response of the graphene. In the static QHE the step structure of the Hall
conductivity is a quantum and topological effect. In the considered case
Eqs. (\ref{grcurry}) does not simply reduce to a topological expression and
the result for the robust plateaus of the nonlinear optical response is
nontrivial. 
\begin{figure}[tbp]
\includegraphics[width=.46\textwidth]{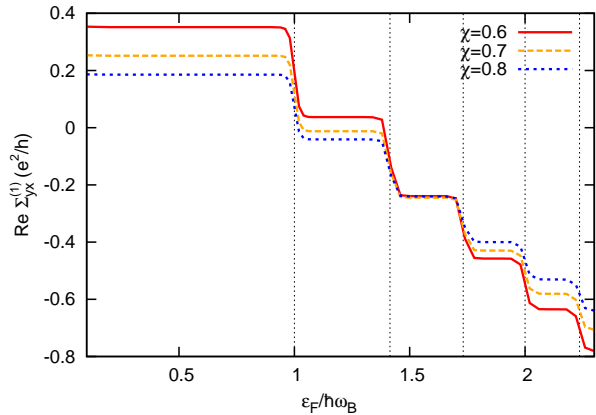}
\caption{Real part of non-linear optical Hall conductivity of graphene QHE
system versus Fermi energy for various pump wave intensities with $\protect%
\omega =0.5\protect\omega _{B}$. The LL broadening is taken to be $\Gamma
=0.2\hbar \protect\omega _{B}$.}
\end{figure}
\begin{figure}[tbp]
\includegraphics[width=.46\textwidth]{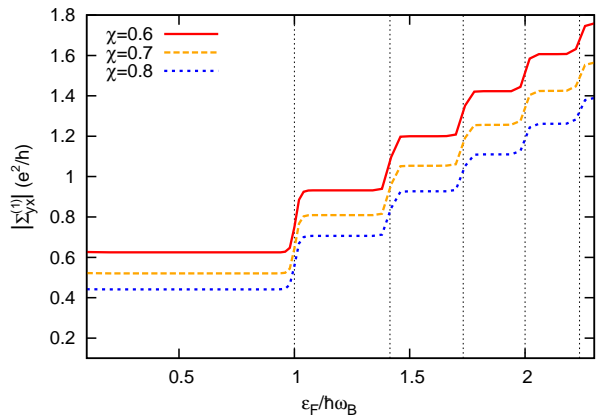}
\caption{Absolute value of non-linear optical Hall conductivity of graphene
QHE system versus Fermi energy for various pump wave intensities with $%
\protect\omega =0.5\protect\omega _{B}$. The LL broadening is taken to be $%
\Gamma =0.2\hbar \protect\omega _{B}$.}
\end{figure}
We further examine how the step-like structure in the nonlinear response of
the graphene behaves for various pump wave frequencies. Absolute value of
non-linear optical Hall conductivity of graphene QHE system at the
fundamental harmonic is shown in Fig. 4. Thus, the step structure preserves
for the wide range of the pump wave frequencies.

We also examine how the step-like structure in the nonlinear optical
response of graphene QHE system behaves as we vary the LL broadening. So we
have calculated $\Sigma _{yx}^{(1)}$ as a function of $\Gamma $, for fixed
values of $\omega $ and $\chi $. We can see from Fig. 5 that, while the
density of states broadens with a width $\sim \Gamma $ the step structure
remains up to large $\Gamma $. 
\begin{figure}[tbp]
\includegraphics[width=.55\textwidth]{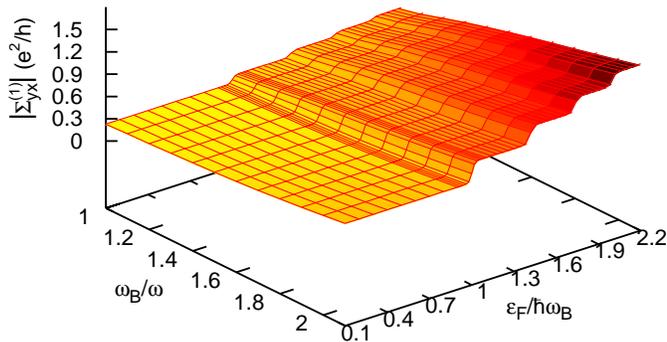}
\caption{Absolute value of non-linear optical Hall conductivity of graphene
QHE system versus Fermi energy and pump wave frequency (ratio $\protect%
\omega _{B}/\protect\omega $) for a wave of intensity $\protect\chi =0.8$.
The LL broadening is taken to be $\Gamma =0.2\hbar \protect\omega _{B}$. }
\end{figure}
\begin{figure}[tbp]
\includegraphics[width=.55\textwidth]{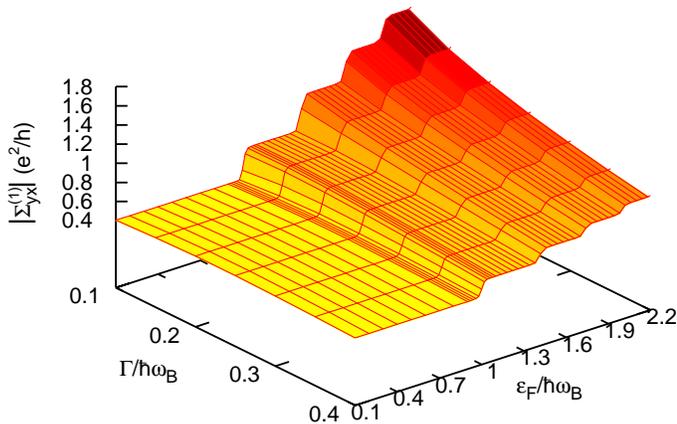}
\caption{Absolute value of non-linear optical Hall conductivity of graphene
QHE system versus Fermi energy and LL broadening for a wave of intensity $%
\protect\chi =0.8$ and frequency $\protect\omega =0.5\protect\omega _{B}$.}
\end{figure}
Note that the non-linear optical Hall conductivity of graphene QHE as in the
case of linear response case defines Faraday rotation and ellipticity of the
transmitted wave. It is straightforward from Eqs. (\ref{F1}) and (\ref{hall}%
) that generated field on the fundamental harmonic polarized perpendicular
to the polarization of a pump wave is defined as 
\begin{equation}
E_{y}^{(g)}=-\frac{2\pi E_{0}}{c}\left( \mathrm{Re}\Sigma _{yx}^{(1)}\cos
\omega t-\mathrm{Im}\Sigma _{yx}^{(1)}\sin \omega t\right) .  \label{Eg1}
\end{equation}%
Hence, the absolute value of Faraday rotation angle ($\vartheta _{F}$) and
ellipticity ($\delta _{F}$) will be defined as: 
\begin{equation}
\left\vert \vartheta _{F}\right\vert =\frac{2\pi }{c}\left\vert \mathrm{Re}%
\Sigma _{yx}^{(1)}\right\vert ;\ \left\vert \delta _{F}\right\vert =\frac{%
2\pi }{c}\left\vert \mathrm{Im}\Sigma _{yx}^{(1)}\right\vert .
\label{Faraday}
\end{equation}%
Thus, Faraday rotation angle and ellipticity present a step-like structure
as the Fermi energy crosses different Landau levels. \ 

\subsection{Third harmonic radiation process}

The nonlinear response of the graphene QHE system to a strong laser
radiation is also proceeded by radiation of harmonics of the incoming
wave-field in the result of the coherent transitions between LLs. As is seen
from Eqs. (\ref{grevol}) and (\ref{grcurry}), the spectrum contains in
general both even and odd harmonics. However, depending on the initial
conditions, in particular, for the equilibrium initial state (\ref{grISPDM})
and at the smooth turn-on-off of the wave field the terms containing even
harmonics cancel each other because of inversion symmetry of the system and
only the odd harmonics are generated. The emission strength of the $s$th
harmonic is characterized by the generalized magneto-optical conductivity of
graphene $\Sigma _{xx}^{(s)}$ and $\Sigma _{yx}^{(s)}$. Calculations show
that for $s>1$ $\left\vert \Sigma _{xx}^{(s)}\right\vert >>\left\vert \Sigma
_{yx}^{(s)}\right\vert $, that is harmonics are radiated with the same
polarization as incoming wave. Hence, we will consider moderately strong
waves and we confine ourselves to only third harmonic radiation process and
numerically investigate generalized longitudinal magneto-optical
conductivity of graphene QHE on the third harmonic $\Sigma _{xx}^{(3)}$. To
determine optimal values of a pump wave frequency for third harmonic
radiation process we calculated $\Sigma _{xx}^{(3)}$ as a function of a pump
wave frequency for various intensities (Fig. 6). As is seen optimal
frequencies are close to the cyclotron frequency $\omega _{B}$.

\begin{figure}[tbp]
\includegraphics[width=.48\textwidth]{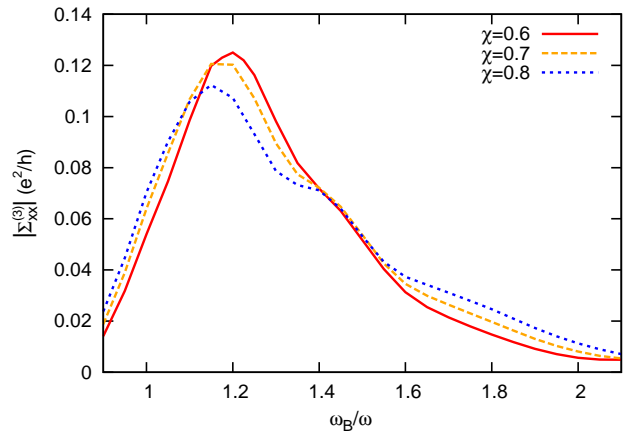}
\caption{Generalized longitudinal magneto-optical conductivity of graphene
QHE on the third harmonic of a pump wave versus ratio $\protect\omega _{B}/%
\protect\omega $ for various intensities at $\protect\varepsilon _{F}/\hbar 
\protect\omega _{B}=0.5$. The LL broadening is taken to be $\Gamma =0.2\hbar 
\protect\omega _{B}$. }
\end{figure}
Then we investigate the variation of conductivity $\Sigma _{xx}^{(3)}$ with
the Fermi energy. In Fig. 7 the third harmonic radiation strength in
graphene QHE system via absolute value of generalized longitudinal
conductivity versus Fermi energy for various pump wave intensities is shown.
From these figure we also notice a step-like structure of the generalized
longitudinal magneto-optical conductivity on the third harmonic. However, $%
\left\vert \Sigma _{xx}^{(3)}\right\vert $ decreases with the increase of
the Fermi energy, which is connected with the fact that due to
nonequidistant Landau levels in graphene the transitions with large energy
difference give main contribution to $\Sigma _{xx}^{(3)}$ for this frequency
range and with the increase of Fermi energy because of Pauli blocking the
contribution of these transitions are vanished. 
\begin{figure}[tbp]
\includegraphics[width=.46\textwidth]{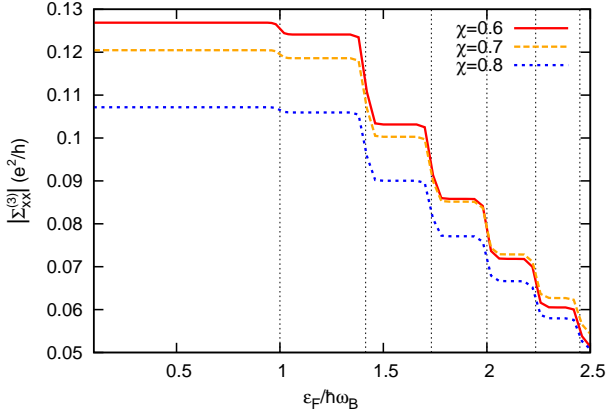}
\caption{The third harmonic radiation strength in graphene QHE system via
absolute value of generalized longitudinal conductivity versus Fermi energy
for various pump wave intensities with $\protect\omega _{B}=1.2\protect%
\omega $. The LL broadening is taken to be $\Gamma =0.2\hbar \protect\omega %
_{B}$. }
\label{eps3}
\end{figure}

We also examine how the step-like structure in the $\Sigma _{xx}^{(3)}$
behaves depending on the pump wave frequency and LL broadening. The results
of our calculations are shown in Figs. (8) and (9). Thus, the step structure
preserves for large LL broadening and for the wide range of the pump wave
frequencies. 
\begin{figure}[tbp]
\includegraphics[width=.46\textwidth]{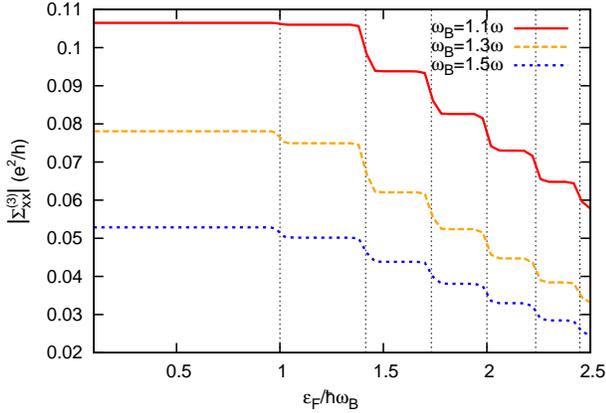}
\caption{The third harmonic radiation strength in graphene QHE system via
absolute value of generalized longitudinal conductivity versus Fermi energy
for various pump wave frequencies with $\protect\chi =0.8$. The LL
broadening is taken to be $\Gamma =0.2\hbar \protect\omega _{B}$. }
\end{figure}
\begin{figure}[tbp]
\includegraphics[width=.46\textwidth]{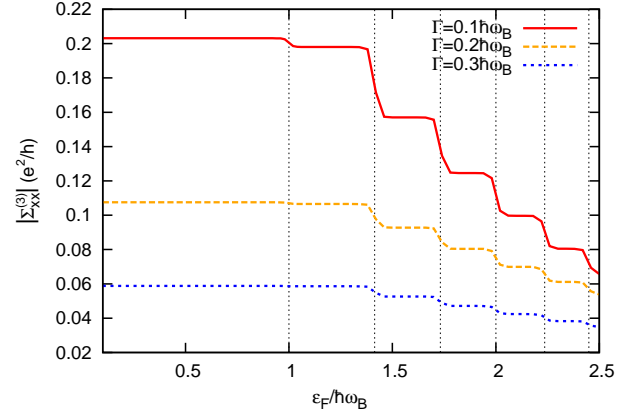}
\caption{The third harmonic radiation strength in graphene QHE system via
absolute value of generalized longitudinal conductivity versus Fermi energy
for various values of LL broadening at fixed pump wave frequency $\protect%
\omega =1.1\protect\omega _{B}$. The intensity parameter is taken to be $%
\protect\chi =0.7$.}
\label{eps7}
\end{figure}

\section{DISCUSSION AND SUMMARY}

Finally let us consider the experimental feasibility of considered
processes. It is clear that in experiment one can observe the considered
effect by measuring $\Sigma _{yx}^{(1)}$ and/or $\Sigma _{xx}^{(3)}$. The
first quantity is responsible for the nonlinear Faraday effect, while last
quantity responsible for third harmonic radiation polarized along the
incoming wave polarization. Thus, the step structure should be observed as
jumps in the intensity of third harmonic ($I_{x}^{(3)}$) or fundamental
harmonic radiation with orthogonal polarization ($I_{y}^{(1)}$). These
quantities are proportional to the pump wave intensity ($I$). From Eqs. (\ref%
{F1}), (\ref{long}), and (\ref{hall})) it follows that $I_{x}^{(3)}=\alpha
^{2}I\left\vert \Sigma _{xx}^{(3)}h/e^{2}\right\vert ^{2}$ and $%
I_{y}^{(1)}=\alpha ^{2}I\left\vert \Sigma _{yx}^{(1)}h/e^{2}\right\vert ^{2}$%
, where $\alpha $ is the fine structure constant. For the pump wave field we
will assume a $\mathrm{CO}_{2}$ laser with $\hbar \omega \simeq 0.1\ \mathrm{%
eV}$. The average intensity of the wave for $\chi =0.8$ is $I\simeq 5\times
10^{7}$ $\mathrm{W/cm}^{2}$. For the setup of Fig. 7 with the chosen
parameters the average intensity of the third harmonic radiation on the
first plateau is $I_{x}^{(3)}\simeq 30$ $\mathrm{W/cm}^{2}$ and the first
step is $\Delta I_{x}^{(3)}\sim 10$ $\mathrm{W/cm}^{2}$. For the setup of
Fig. 4 the average intensity on the first plateau is $I_{y}^{(1)}\simeq 180$ 
$\mathrm{W/cm}^{2}$ and the first step is $\Delta I_{y}^{(1)}\sim 280$ $%
\mathrm{W/cm}^{2}$. Thus, that the present effect is well within the
experimental feasibility.

To summarize, we have presented a microscopic theory of the graphene
interaction with coherent electromagnetic radiation in the quantum Hall
regime. The evolutionary equation for a single-particle density matrix has
been solved numerically. We have revealed that the nonlinear optical
response of graphene to a strong laser radiation in the quantum Hall regime,
in particular, radiation strength on the third harmonics, as well as
nonlinear Faraday effect, has a characteristic Hall plateau structures that
persist for a wide range of the pump wave frequencies and intensities even
for significant broadening of \ LLs because of impurities in graphene.

\begin{acknowledgments}
This work was supported by the RA MES State Committee of Science, in the
frames of the research project No. 15T-1C013.
\end{acknowledgments}

\end{document}